\documentstyle[12pt,openbib,psfig]{article}
\newcommand{\f}{\frac}

\newcommand{\D}{\Delta}

\newcommand{\be}{\beta}

\newcommand{\om}{\omega}
\newcommand{\ep}{\epsilon}

\def\be{\begin{equation}}
\def\ee{\end{equation}}

\begin{document}

\title{\bf Have we observed the skin vibration of realistic strange stars (ReSS) ?}

\author{ Monika Sinha $^{1,~2,~3}$,  Jishnu Dey $^{2,~ 4~ \dagger}$,
Mira Dey $^{1,~ 4 ~\dagger}$,\\ Subharthi Ray $^{5}$ and
Siddhartha Bhowmick, $^{6}$ }

\maketitle

\begin{abstract}
{Skin vibration of ReSS and consequent resonance absorption can
account for the absorption lines in the spectrum of X-ray emission
from many compact stellar objects and in particular, the stars
J1210$-$5226 and RXJ1856$-$3754. Observations of the X-ray
spectrum of these stars is difficult to explain, if they are
neutron stars. }

\end{abstract}

\vskip .2cm

keywords: compact stars~--~realistic strange stars~--~dense
matter.

\vskip .2cm

\noindent $^1$ Dept. of Physics, Presidency College, 86/1 College
Street, Kolkata 700 073, India\\ $^2$ Azad Physics Centre, Dept.
of Physics, Maulana Azad College, 8  Rafi Ahmed Kidwai Road,
Kolkata 700 013, India\\ $^3$ CSIR-NET fellow, Govt. of India\\
$^4$ Senior Associate, IUCAA, Pune, India \\ $^5$ FAPERJ Fellow,
Instituto di Fisica, Universidade Federal Fluminense, Niteroi, RJ,
Brasil \\ $^6$ Department of Physics, Barasat Govt. College,
Barasat, North 24 Parganas, W. Bengal, India\\ $\dagger$ permanent
address; 1/10 Prince Golam Md. Road, Kolkata 700 026,\\ India;
e-mail : deyjm@giascl01.vsnl.net.in.\\ $^*$ Work supported  in
part  by DST  grant no. SP/S2/K-03/2001, Govt. of India.

\section{Introduction}

The spacecraft observations of X-ray emission from compact
objects, specially the outstanding capabilities of the
{\it{Chandra}} X-ray observatory have greatly increased our
potential to study the characteristics of the surface radiation
from compact stars.

According to Pavlov et al. \cite{pav} (PZS in short) the
best-investigated compact central object J1210$-$5226 in the
supernova remnants of G296.5+10 shows two absorption lines of 0.7
keV and 1.4 keV which are equally strong. Attempts to explain the
absorption features as caused by the intervening interstellar (or
circumstellar) medium material lead to huge overabundance for some
elements. Therefore the observed lines are most likely intrinsic
to the compact star.

The star RXJ1856.-3754 is a compact object with no pulsation and
is only about 120 pc away from us. There is a recent controversy
\cite{pons, drake} about this star since it shows a black body
spectrum but with a dip at about 0.3 keV which has led to some
speculation about X-ray activity comprising of  two zones (with
their associated magnetic fields) and two temperatures.

In the present paper we show that a natural explanation of the
dip(s) is in terms of absorption at the resonance frequency
(fundamental or its harmonics) of the star surface which is only
$10^{-7}~ fm$ thick. For this we employ the EOS1, \cite{d98} (SS1
in \cite{Li99a, Li99b}) which has been widely used by many authors
for explaining (1) the Mass-Radius (M-R) curves of stars
\cite{d98,Li99a}, (2) properties of quasi-periodic oscillations
\cite{Li99b}, (3) continuity of the entropy from the hadron to
quark phase \cite{aa},  (4) fast rotation properties of stars
\cite{dorota, tbd}, (5) gamma ray bursts \cite{bd,qnova}, (6)
radial oscillations \cite{sharma, mpla} (7) superbursts and long
bursts through quark pairing as surface phenomenon of compact
objects \cite{sinha} - and many other properties. The pressure
derivative at the star surface is known to be related to the
vibrations of the skin, which in our case consists of the lighter
quarks with mass $\sim$ 130 MeV (we neglect strange quarks which
have mass exceeding the u,d-mass by another 150 MeV). This gives
us a natural frequency for the system and any vibration at the
surface will be preferentially absorbed at this frequency
according to the basic tends of physics. Therefore the system will
absorb this frequency and its harmonics preferentially from the
X-ray generated at the surface by accreting matter falling in from
outside. This is our model for absorption band in the spectra.

As a by-product we also calculate properties of the electron cloud
associated with the star and electron natural frequency controlled
by the central positive charge.

We are grateful to the referee for pointing out that one could
assume that the absorption lines are due to once - ionized helium
in a super strong magnetic field. If one assumes these are
electron cyclotron lines, the features might be interpreted as the
fundamental and the first  harmonic of the electron cyclotron
energy in a magnetic field as two fundamentals from two regions
with different magnetic fields. However PZS clearly point out that
this is difficult to believe on two counts:

\begin{enumerate}
\item The surface magnetic field has to be $B~>~3\times10^{12}~G$
\item The oscillator strength of the first harmonic is smaller
than the fundamental by a factor 0.002 so that it is hard to
explain why the 1.4 keV  is as strong as the 0.7 keV feature if we
assume that the two lines are associated with the same magnetic
field.
\end{enumerate}

The referee has also asked us to comment on the recent paper by
Cottam, Paerels and Mendez \cite{CPM}where they have found the red
shift of a star to be 0.35 and assume that `for astrophysically
plausible range of masses  $(M\sim ~ 1.3 - 2.0 ~ M_\odot$ it
excludes some models in which the neutron stars are made of more
exotic matter' including model of \cite{d98}. While this may be
so, there may be other stars whose red shift may be measured in
the near future which will be shown to be compatible with the
latter model. More importantly Li et al.{Li99b} have pointed out
that the star 4U 1728$-$34 has mass less than 1.1 $M_\odot$ so
that we see no reason why such a mass is not astrophysically
plausible, if a star emitting X rays. In fact the crucial proof
for the existence of strange stars may come from the pinpointed
low mass compact object of mass 1.1 $M_\odot$ or less, since low
masses with neutron matter EOS cannot form stars. Strange quark
stars are self bound, so they can have low masses and if they are
fast rotating with binary partners they may also show up
marginally stable orbits due to their oblate structure and can
lead to observable phenomenon as found by Zdunik and Gourgoulhon
\cite{ZG}.

\section{ Incompressibility and oscillations in ReSS. }

Incompressibility K, plays a very significant role in nuclear
physics since the energy per particle, E/A has a minimum at normal
density and at zero temperature. The pressure, p is zero at this
point. For such systems the energy can be written out as the sum
of the equilibrium value and a term proportional to the square of
the change in the radius, leading to the spectrum of a harmonic
oscillator. It was found in \cite{bdp} that the magnitude of
nuclear incompressibility is comparable to that of a nucleon and
of quark gas. In \cite{mulders} it was found that the excitations
of the nucleon could be related to K and the bag inertia
parameter.

The rate of change of p with number density is related to the
incompressibility and the breathing modes determined by it. Since
the ReSS strange matter shows a minimum at about five times normal
density, the star surface has this density and this surface will
vibrate. The vibration frequency spectrum is controlled by
$dp/dr$.

\begin{figure}[htbp]
\centerline{\psfig{figure=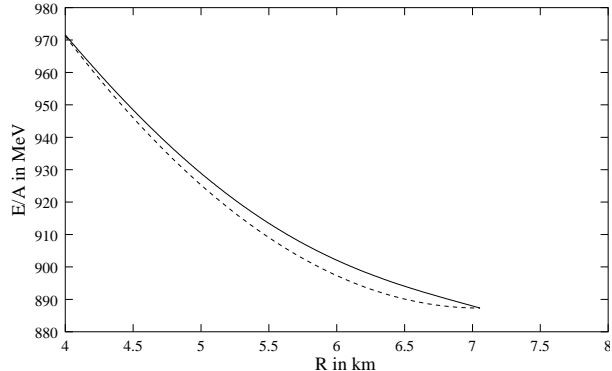,width=8cm}}
\caption{\footnotesize{Energy per baryon {\it vs.} radius.}}
\label{min}
\end{figure}

In ReSS model the energy per particle has a minimum at the
surface. The nature of the curve near the minimum can be
approximated by a harmonic oscillator as shown in Fig. \ref{min}.
A Taylor expansion of the energy about $r=R$ gives \be \f{E(r)}A =
\f{E(R)}A + \f12k(r)~~ (r-R)^2 \ee

where R is the star radius, $k(r)~=~-4\pi~r^2\f{dp}{dr}$ and \be
\f{dp}{dr}=\f{-G(p(r)+\ep)(m(r)+4\pi~r^3p(r))}{r^2(1-\f{2Gm(r)}{r})}
\ee is the TOV equation with conventional notation. The frequency
of the vibration is given by \be \om~=~\sqrt\f{k(R)}{m_{skin}}.
\ee We take \be m_{skin}~=~ 4\pi R^2~n(R)\f{m_u(R)+m_d(R)}{2}d \ee
where $n(R)$ is the number density at the surface, while $m_u$ and
$m_d$ are the respective u and d masses obtained from the self
consistent ReSS solutions \cite{d98}. To estimate the skin depth d
we turn to the electron cloud outside the star.

  It is well known \cite{alcock,usov} that strong (or
gravitational interaction alone) cannot retain the electrons
within the sharp surface of a strange star and they spread outside
the star in a shell of thickness several hundred fm \cite{usov2}.
They induce a positive charge at the centre once the charge
neutrality condition is set up over the years and the
electrostatic potential holds them. The minimum number of
electrons in the shell that ensures this equilibrium is \be
N_{crit} ~=~ \f{137~R}{\hbar c} (\sqrt{(k_f(R)\hbar
c)^2+m_e^2~c^4}-m_e~c^2), \ee where $\hbar c~ k_f(R)$, the
electron Fermi momentum is $31.28$ MeV at the surface and $m_e$ is
the mass of the electron.

Although $N_{crit}$ is large, $1.507\times 10^{20}$, it is
negligibly small compared to the number $N$ that can be present
even in an infinitesimally small skin depth. Enhance the
electrostatic attraction by a factor $f \sim 10$, with the total
number of electrons in the shell is $f N_{crit}$ : since number
$\sim~{k_f}^3$, an infinitesimal change in the Fermi surface by an
amount $\D k_f(R)\sim~10^{-8}~ fm ^{-1}$ accounts for $N_{crit}$
as \be \f{3~\D k_f(R)}{k_f(R)}~~=~~ \f{N_{crit}f}{N}~\ee. For a
skin depth  $d~=~1.5\times 10^{-7}~fm$, the total number of
electrons is \be N~=~4\pi R^2 \f{k_f(R)^3}{3\pi^2}d ~\approx~
10^{28}. \ee

Different central densities lead to different stellar masses and
radii. This leads to variations in the fundamental frequency of
skin vibrations. We find that resonance absorption for the first
and the second harmonics of a star with mass $1.22~ M_\odot$ and
radius $7.16 ~km$ corresponds to the $0.7$ and $1.4$ $keV$
absorption bands of the star J1210$-$5226, the puzzling pulsar in
G296.5$-$10 (PZS). The $0.3~keV$ absorption band of RX
J1856.5$-$3754 corresponds to the fundamental of a strange star
with mass $0.94~ M_\odot$ and radius $6.74 ~km$ (PZS).

The factor f can, in principle be measured from the electron
oscillation in the outside cloud that gives out a radio wave of
energy \be h \nu \le \hbar \sqrt{\f{2 f N_{crit} e^2}{m_e R^3}}.
\ee This may be a broad band radio emission since compressional
modes are known to lead to giant absorption (or emission, as the
case may be) bands in nuclei and furthermore the electrons in the
cloud are constantly moving leading to considerable extra
broadening of the band. The referee kindly points out to us that
such bands are indeed present in radio observations.

For the strange star fitting J1210$-$5226, $f = 10$ gives a radio
wave with wavelength more than 180 meters with possible harmonics
in the decameter radio band.

\begin{table}[htbp]
\caption{Resonant frequency of stars} \vskip 1cm
\begin{center}
\begin{tabular}{|c|c|c|c|c|}
\hline $\rho_c~10^{14}$&$M/M_\odot$&R&$\nu~10^{16}$&$e_\nu$ \\
g/c.c.&&km&/sec.&$keV$\\ \hline 14.85&0.407&5.262&5.3283&0.220\\
15.35&0.502&5.613&5.6857&0.235\\ 15.85&0.607&5.941&6.0606&0.251\\
16.35&0.719&6.247&6.4533&0.267\\ 17.35&0.841&6.536&6.8802&0.285\\
17.85&0.894&6.644&7.0635&0.292\\ 18.35&0.943&6.741&7.2455&0.300\\
19.35&1.037&6.970&7.6103&0.315\\ 21.35&1.159&7.051&8.1407&0.337\\
22.35&1.205&7.139&8.3663&0.346\\ 22.85&1.226&7.162&8.4758&0.350\\
24.85&1.288&7.214&8.8199&0.365\\ 25.35&1.300&7.228&8.8954&0.368\\
26.85&1.333&7.237&9.1107&0.377\\ 28.35&1.357&7.240&9.2834&0.384\\
28.85&1.363&7.240&9.3365&0.386\\ 30.85&1.386&7.232&9.5348&0.394\\
32.85&1.402&7.218&9.6950&0.401\\ 34.85&1.414&7.199&9.8368&0.407\\
36.85&1.423&7.178&9.9558&0.412\\ 38.85&1.429&7.154&10.0608&0.416\\
40.85&1.443&7.130&10.1516&0.420\\
42.85&1.435&7.105&10.2311&0.423\\
44.85&1.435&7.105&10.3012&0.426\\
46.85&1.437&7.055&10.3620&0.429\\ \hline
\end{tabular}
\end{center}
\label{res}
\end{table}

\begin{figure}[htbp]
\centerline{\psfig{figure=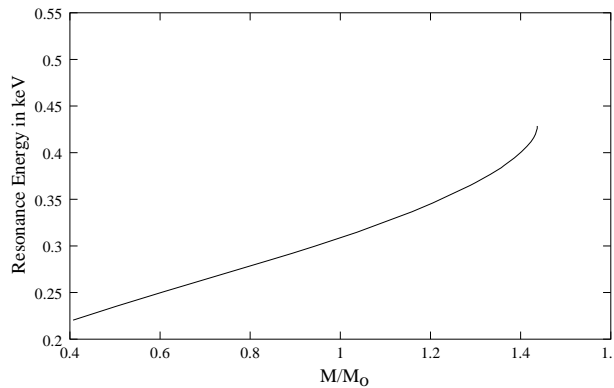,width=8cm}}
\caption{\footnotesize{Resonant energy (fundamental) for skin
vibrations of strange stars with different masses.}} \label{resen}
\end{figure}

In summary, we have pointed out the exciting possibility that the
absorption bands seen with Chandra and XMM-Newton in thermal X-ray
spectra of stars could arise from surface vibrations of the u-d
quarks on the star surface, if the stars are composed of strange
matter.

We hope in future, small stars with mass less than or roughly
equal to that of the sun will be found from data and their radius
found to be of the order of $\sim 5 $ to $7$ km. This would lead
to a rich interplay between X-ray astrophysics and QCD.

The authors are grateful to Dr. Alak Ray of TIFR for pointing out
the important paper PZS to them. It is a pleasure to thank Dr.
Rajesh Gopakumar of HRI for pointing out the importance of $\D
k_f$. MS, JD and SB thank IUCAA for a short pleasant stay. We are
grateful to the anonymous referee for improving our paper
considerably.



\end{document}